\documentclass[numberedappendix]{emulateapj}
\begin{document}

\slugcomment{Accepted by ApJ: April 17, 2007}

\title{
An Inner Hole in the Disk around TW Hydrae 
Resolved in 7~millimeter Dust Emission}

\author{
A. M. Hughes \altaffilmark{1},
D. J. Wilner \altaffilmark{1},
N. Calvet\altaffilmark{2},
P. D'Alessio\altaffilmark{3},
M. J. Claussen \altaffilmark{4},
M. R. Hogerheijde \altaffilmark{5}
}

\email{mhughes@cfa.harvard.edu}

\altaffiltext{1}{Harvard-Smithsonian Center for Astrophysics,
  60 Garden Street, Cambridge, MA 02138}
\altaffiltext{2}{Department of Astronomy, University of Michigan, 
  500 Church Street, Ann Arbor, MI 48109}
\altaffiltext{3}{Centro de Radioastronom\'{i}a y Astrof\'{i}sica, UNAM, 
  Apartado Postal 3-72 (Xangari) 58089 Morelia, Michoac\'{a}n, Mexico}
\altaffiltext{4}{National Radio Astronomy Observatory, P.O. Box O,
  Socorro, NM 87801}
\altaffiltext{5}{Leiden Observatory, Leiden University, P.O. Box 9513, 2300 RA, 
  Leiden, The Netherlands}


\bibliographystyle{apj}

\begin{abstract}
We present Very Large Array observations at 7~millimeters wavelength
that resolve the dust emission structure in the disk around the
young star TW Hydrae at the scale of the $\sim$4~AU (~0\farcs16) radius 
inner hole inferred from spectral energy distribution modeling. 
These high resolution data confirm directly the presence of an 
inner hole in the dust disk and reveal a high brightness ring that 
we associate with the directly illuminated inner edge of the disk. 
The clearing of the inner disk plausibly results from the 
dynamical effects of a giant planet in formation. 
In an appendix, we develop an analytical framework for the interpretation 
of visibility curves from power-law disk models with inner holes.
\end{abstract}

\keywords{circumstellar matter ---
planetary systems: protoplanetary disks ---
stars: individual (TW~Hydrae)}

\section{Introduction}
The TW Hya system is thought to be a close analog of the early 
Solar nebula.  At a distance of 51$\pm$4 pc \citep{mam05}, it is 
the closest known classical T Tauri star, and a suite of observational 
studies have shown that TW Hya harbors a massive disk of gas and dust.
Scattered light observations at optical and near-infrared wavelengths
reveal a surface brightness profile consistent with a nearly face-on, 
optically thick, flared disk extending to $\sim$200~AU in radius 
\citep{rob05,wei02,kri00,tri01}.
Observations at millimeter wavelengths have detected thermal dust 
emission and a variety of molecular species, including $^{13}$CO, 
$^{12}$CO, CN, HCN, HCO$^+$, and DCO$^+$ \citep{wei89,zuc95,kas97,dis03,
wil03,qi04}.
The dust also displays signatures of grain growth up to centimeter 
scales \citep{wil05}, and perhaps substantially larger sizes.

Detailed models of the TW Hya spectral energy distribution (SED) 
provide constraints on many aspects of the disk structure \citep{cal02},
including the radial dependence of outer disk surface density and 
temperature, and a clearing of the inner disk within $\sim$4~AU radius. 
Resolved interferometric observations of millimeter and submillimeter 
dust emission are in good agreement with the structure inferred from the 
irradiated accretion disks models that match the SED \citep{qi04,wil00}, 
though the resolution and sensitivity at these wavelengths have not been 
sufficient to address the presence of the inner hole. 

The inner hole is indicated by two features of the SED \citep{cal02}: 
(1) a flux deficit from $\sim$2-20$\mu$m, indicative of low (dust) 
surface density in the inner disk, and 
(2) a flux excess at $\sim$20-60 $\mu$m, thought to originate from 
the truncated inner edge of the disk, directly illuminated by the star.  
Similar spectral features have been recognized in other T Tauri star SEDs
\citep[e.g. GM Aurigae and DM Tauri; see][]{cal05} and may signify 
an important phase in the evolution of circumstellar disks.
One exciting possibility is that a discontinuity in the inner disk is a 
consequence of the perturbative gravity field of a giant planet.
Theories of planet-disk interaction predict the opening of gaps in a disk 
as a result of the formation of massive planets \citep[e.g.][]{lin86,bry99}.
However, \citet{bos93,bos96} show that the interpretation of infrared
flux deficits as central clearings is not unique and reproduce SEDs of 
accreting disks around low-mass, pre-main sequence stars with a
combination of opacity and geometry effects in the unresolved system.
Spatially resolved observations of disk structure are required to confirm 
the inference from spectral deficits of inner disk clearing.  

To probe the disk morphology on size scales commensurate with the 
4~AU transitional radius of \citet{cal02}, we have used the 
Very Large Array\footnote{
The National Radio Astronomy Observatory is a facility of the National 
Science Foundation operated under cooperative agreement by Associated 
Universities, Inc.} to observe thermal dust emission from TW Hya at
a wavelength of 7~millimeters. These observations show clearly a deficit 
of dust emission in the inner disk consistent with the predicted hole.

\section{Observations}

\begin{figure*}
\centering
\includegraphics[scale=0.6,angle=90]{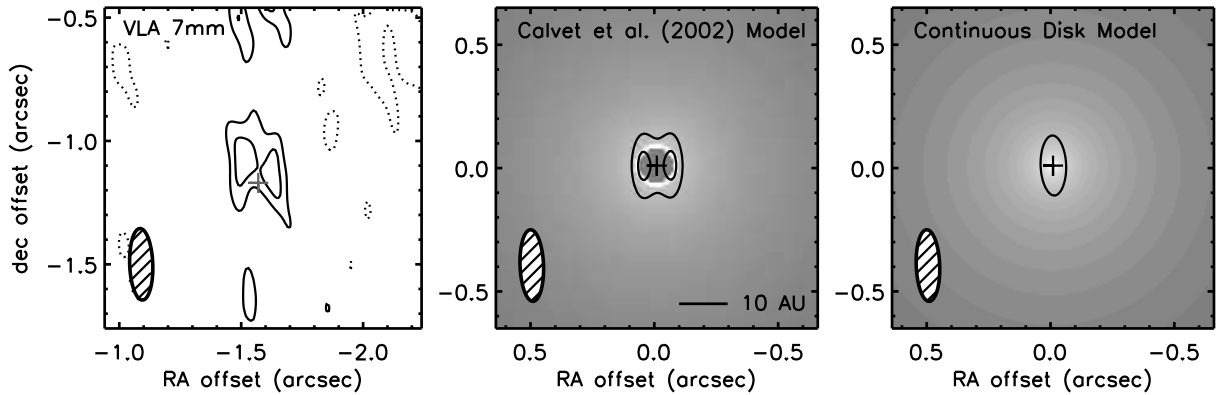}
\caption{
The TW Hya 7 millimeter continuum emission observed with the VLA {\em (left)} 
compared to simulated images generated from the \citet{cal02} model of an 
irradiated accretion disk, truncated at the 4~AU radius indicated 
by the SED and including the directly illuminated inner edge {\em (center)},
or extending in to the 0.01~AU dust destruction radius {\em (right)}. The 
contour levels are $-2,2,3\times$0.23~mJy (the rms noise).  In each panel, 
the ellipse in the lower left corner indicates the $0\farcs29 \times 
0\farcs09$, PA $2.1^\circ$ synthesized beam. The cross marks the derived 
position of the central star (see text).
The center and right panels also show the disk models at full resolution, in 
a logarithmic grayscale to display the range of intensities in the fainter, 
outer regions of the disk and the bright, thin wall at the inner edge of
the canonical model.
}
\label{fig:map}
\end{figure*}

We used the Very Large Array to observe TW Hya at 7 millimeters in 
the most extended (A) configuration.
The observations used 23 VLA antennas (several were unavailable due to 
eVLA upgrades) that gave baseline lengths from 130 to 5200 k$\lambda$. 
The observations were conducted for four hours per night on
10, 11 February 2006 and 7 March 2006, from 7 and 11 UT (0 to 4 MST), 
during the late night when atmospheric phases on long baselines are 
most likely to be stable. Both circular polarizations and two 50 MHz 
wide bands were used to obtain maximum continuum sensitivity.
The calibrator J1037-295 was used to calibrate the complex gains,
using an 80-second fast switching cycle with TW Hya.
The calibrator J1103-328, closer to TW Hya in the sky, was also 
included in a few minutes of fast switching each hour to test of the 
effectiveness of the phase transfer from J1037-295.  
The phase stability was good during the observations of 11 February, 
worse on 10 February, and much worse on 7 March.
Using the AIPS task SNFLG, we pruned the data with phase jumps of more 
than 70$^\circ$ between phase calibrator scans. This procedure passed 
about 80\% of the data from the night of 11 February but substantially
less from the other nights. Therefore, in the subsequent analysis
we have used data only from 11 February, the most stable night.  
The calibrator 3C286 was used to set the absolute flux scale (adopting 
1.45 Jy, from the AIPS routine SETJY), and we derived 1.95 Jy for 
J1037-295. The uncertainty in the flux scale should be less than 10\%.

\section{Results}

\subsection{7~mm Image}
Figure~\ref{fig:map} shows an image of the TW Hya 7 millimeter emission, 
where a Gaussian taper (FWHM 2000~k$\lambda$) has been used in the imaging
process to obtain an angular resolution matched to the surface brightness 
sensitivity.  The rms noise level in this image is 0.23 mJy, and the peak flux 
of 0.88 mJy corresponds to a brightness temperature of 46 K.  Because of the 
southern declination of the TW Hya and the VLA antenna geometry, the 
beam is elliptical and the resolution is higher east-west than north-south.
Inspection of Figure~\ref{fig:map} shows that the 7~millimeter emission is 
clearly not centrally peaked, as would be expected for a disk that extends 
continuously inwards towards the central star.  Instead, the image exhibits 
a double-peaked morphology, consistent with a nearly face-on disk with 
a central hole observed with an elliptical beam. An image of the test
calibrator J1103-328 made with the same parameters is point-like, as expected. 

\subsection{Radially Averaged 7~mm Visibilities}
\label{sec:deproj}
The central hole in the TW Hya emission may be identified even more 
clearly in the visibility domain, free from the effects of the Fourier 
transform process and non-linear deconvolution.
To better show the brightness distribution, we averaged the visibility 
data in concentric annuli of deprojected (u,v) distance, ${\cal R}_{uv}$, 
as described in \citet{lay97}. We use the TW Hya disk position angle 
and inclination found by \citet{qi04} from CO line imaging of the outer disk
(-45$^\circ$ and 7$^\circ$, respectively). These values may not be valid in 
the inner disk, if e.g. the disk warps in the interior. However, as long as 
the disk remains close to face-on at all radii, the deprojection correction 
is small and therefore insensitive to the exact values of these parameters. 

For each visibility, the coordinates were redefined in terms of 
${\cal R}=\sqrt{u^2+v^2}$, the distance from the origin of the (u,v) plane, 
and $\phi=\arctan{\left(\frac{v}{u}-PA\right)}$, the polar angle from 
the major axis of the disk (defined by the position angle, $PA$, 
measured east of north). Assuming circular symmetry and taking into 
account the disk inclination $i$, the deprojected (u,v) distances 
parallel to the major and minor axes of the disk, are  
$d_a = {\cal R} \sin{\phi} $, $d_b = {\cal R} \cos{\phi}~\cos{i} $, 
respectively, and the deprojected (u,v) distance is 
${\cal R}_{uv} = \sqrt{ d_a^2 + d_b^2}$.

An important parameter in the averaging process is the position of the star, 
or the center of the disk.  
To examine the radial distribution of flux at the smallest scales permitted
by the data, particularly in the east-west direction of highest resolution,  
we must know the phase center to within a fraction of the radius 
set by the resolution, i.e., the position of the star must be specified to 
within a few hundredths of an arcsecond, which is better than the absolute astrometric accuracy of the 
data (typical positional accuracies in the A array are $\sim0\farcs1$ due to 
baseline uncertainties and uncorrected tropospheric phase fluctuations; see, 
e.g. the 2004 VLA Observational Status Summary). 
To the extent that the 
disk is symmetric, the process of deprojecting and averaging at the correct 
star position will minimize the scatter within each deprojected radial bin and 
bring the average of the imaginary parts of the visibilities (the average 
phase) to zero. Therefore, we chose the star position to be that which 
minimized the absolute value of the mean of the imaginary visibility bins. 
This position is indicated by the cross in the left panel of Figure~\ref{fig:map}.  

\begin{figure}
\centering
\plotone{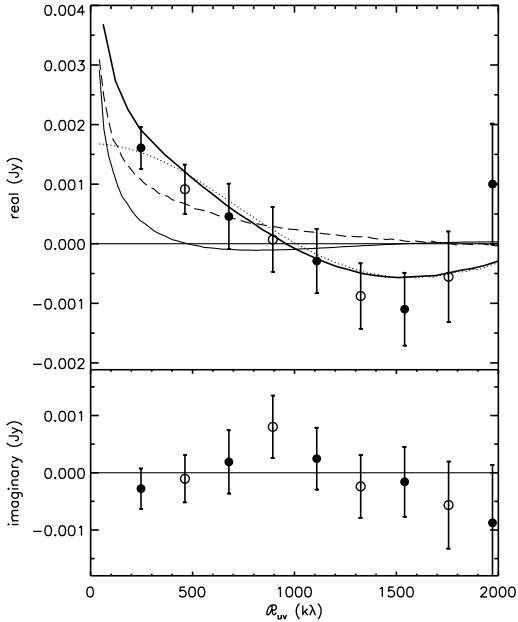}
\caption{TW Hya real and imaginary 7 millimeter visibilities, in 
430-k$\lambda$ bins of deprojected u-v distance from the center of the disk.
Error bars represent the standard error of the mean in each bin.  
Note that bins represented by open and filled circles are not independent: 
the bins are overlapping, i.e., the domain of each open circle extends 
to the neighboring filled circles.  
The calculated visibility functions for the best fit irradiated accretion 
disk model is indicated by the heavy solid line and is composed of 
contributions from an outer disk (solid line) with an inner hole of 
radius 4.5~AU and a bright, thin wall (dotted line). 
For comparison, the visibility function for a disk model that extends in to 
the dust destruction radius at $\sim$0.01~AU is also shown (dashed line).
To the extent that the disk is radially symmetric, the imaginary part of the
visibility (the average phase) should be zero for all models.  
}
\label{fig:vis}
\end{figure}

Figure~\ref{fig:vis} shows the annularly averaged visibility amplitude as 
a function of ${\cal R}_{uv}$. The width of each bin is 430~k$\lambda$, chosen 
to be narrow enough to sample the shape of the visibility function 
and also wide enough to have sufficient signal-to-noise ratio. 
Although the visibility data are still noisy when divided up in this 
way, it is evident that the visibility function passes through a null near 
${\cal R}_{uv}$ of $\sim$1000~k$\lambda$, indicative of a sharp edge in the emission.

\section{Discussion}
The most striking feature of the new 7~millimeter observations is 
the central depression in the image, which is also indicated by the 
presence of the null in the deprojected visibility function. 
We identify this feature with a clearing of the inner dust disk.
A continuous disk that extends inward to the dust destruction 
radius at $\sim$0.01~AU would show sharply centrally peaked emission 
and would not show a null at the observed baselines. 
Figure~\ref{fig:map} compares the 7~millimeter image (left panel) with 
an image generated for a model with a continuous disk (right panel). 
Figure~\ref{fig:vis} also shows the visibility function of continuous disk 
(dashed line); this is clearly not compatible with the 7~mm observations 
(or the infrared SED ).

Independent of detailed modeling, the size scale of this inner hole 
can be estimated simply from the separation of the peaks in the image.
These peaks are sensitive primarily to inner edge of the disk, 
where the 7~millimeter brightness is highest. The separation of the
peaks is $\sim0\farcs14$, or $\sim7$~AU.
This separation is slightly smaller than the diameter of the inner hole, 
since the bright emission from the inner edge to the north and 
to the south of the star, combined with the lower angular resolution 
in the north-south direction, tend to draw the image peaks together. 
This effect is evident 
in the right panel of Figure~\ref{fig:map}, in which the contoured 
peaks can be seen to lie interior to the inner edge of the disk.
The null in the visibility function provides a corroborating estimate
of the size of the inner hole in the disk.
In appendix~\ref{sec:viscurves}, we show how the angular scale of 
the null in the visibility function of a power law disk depends on 
the density and temperature power law indices and the radius of 
the inner hole. 
If we assume that the outer disk emission contributes little on these 
long baselines, and that the total emission is dominated by a bright, 
thin ring associated with the inner edge of the disk, 
as discussed in \S\ref{sec:analytic} below, then we can estimate the 
radius of the inner hole hole with equation~\ref{eq:bessel}. 
A linear fit to the binned visibilities gives a null position of
930$\pm$60 k$\lambda$ and implies an inner hole radius 4.3$\pm$0.3~AU.  
Thus the resolved dust emission shows an inner hole in the disk at 
a size scale very similar to that inferred from SED modeling.

\subsection{Comparison with Disk Models}

\subsubsection{Power Law Disk Models}
\label{sec:analytic}

The dust emission from the outer disk of TW Hya has been shown to be
a good match to the structure of an irradiated accretion disk model,
approximated by power laws in temperature and surface density with 
indices $q\sim0.5$ and $p\sim1.0$, respectively, over a wide range of radii.
However, an extrapolation of this very simple power law model to an 
inner disk truncated at $\sim4$~AU radius is not compatible with the new 
long-baseline 7~millimeter data.  For these power laws, the null in the 
visibility function of the 4~AU hole should appear at $\sim500$~k$\lambda$, 
which is a factor of two smaller than observed (equation~\ref{eq:approx}). 
In order for a 4~AU hole to be consistent with a $\sim$1000~k$\lambda$ null, 
the power law model requires a much steeper emission gradient, with the sum 
of the radial power law indices $p+q$ approaching $5$. Such steep power laws 
are inconsistent with observations of the outer disk.  In addition, this 
power law model fails to reproduce the flux in the image peaks by nearly 
an order of magnitude.  Another possibility is that the radius of the hole 
is smaller than that predicted by the SED.  A power law disk with $p+q = 1.5$ 
and radius 2~AU does reproduce the 1000~k$\lambda$ null and observed peak 
separation; however, this model still fails to reproduce the observed 
peak flux by more than a factor of five.  These discrepancies indicate 
that a more complex model of the disk is needed.  In general, reproducing 
the observed peak flux in the new high resolution observations requires a 
greater concentration of material at the inner edge of the disk than that of 
power-law disk models.  

A natural modification to the power law disk model with an inner hole 
is the addition of a bright, thin, inner edge, or ``wall'' component.
This ``wall'' component corresponds to the frontally illuminated 
inner edge of the disk in the calculations of \citet{cal02}, who show 
that a small range of temperatures is required to reproduce the narrow
spectral width of the mid-infrared excess.
The presence of this additional compact component to the model shifts the 
angular scale of the null in the visibility function to larger ${\cal R}_{uv}$ 
and raises the flux at long baselines. In this composite model, for a given 
power law description, the angular scale of the null in the composite 
``disk+wall'' visibility function depends on (1) the radius of the inner hole, 
and (2) the relative brightness of the disk and wall.
The effect of the second dependency is to move the angular scale of the 
null between the 
limiting positions from the ``disk'' alone and from the `` wall'' alone.  
In fact, the position of the null in the 7~millimeter data is close to 
that expected from an infinitesimally thin ring of $\sim$4~AU radius 
(equation~\ref{eq:bessel}). A bright, thin ring also reproduces the 
$\sim 0\farcs14$ separation of the image peaks. 
Thus it appears that the high resolution observations show primarily 
the directly illuminated wall at the inner edge of the disk. At these 
long baselines, the extended emission from the outer disk that dominates 
at larger size scales is weak and effectively not detected. 

\subsubsection{Irradiated Accretion Disk Model}
\label{sec:ratran}

The irradiated accretion disk model of \citet{cal02} provides a more 
realistic description of disk structure than a simple power-law model.
To compare the data to this more sophisticated disk model,
we simulate numerically the 
expected emission at 7~millimeters, including the detailed visibility 
sampling of the Very Large Array observations. 

For the frontally illuminated component of
the inner edge of the disk, we follow the prescription of \citet{cal02}, 
adopting a shape given by 
\begin{eqnarray}
z_s = z_0 \exp{[(R-R_0)/\Delta R]}
\end{eqnarray}
and temperature
\begin{eqnarray}
T_{phot}(R) \approx T_* \left( \frac{R_*}{R} \right)^{1/2} 
                        \left( \frac{\mu_0}{2} \right)^{1/4}
\label{eq:temp}
\end{eqnarray}
where $R$ is the radius from the star, $z_s$ is the height of the wall,
subscript $0$ refers to the boundary between the wall and the outer disk,
$\mu_0 = \cos{\theta_0} $ and $\theta_0 = \pi / 2 - \tan^{-1}(dz_s/dR)$,
and we assume that at 7~millimeters the radial brightness tracks the 
temperature. To set the absolute flux of this component,
we normalized the intensity distribution to match 
the peak flux of the image and the total flux of the disk determined from 
previous, lower resolution imaging \citep{wil00}. 
Since $\Delta R$ is not well constrained by our data, except that the wall 
must be narrow compared to the size of the hole ($\Delta R \ll R$) and the 
resolution of the data ($\Delta R \ll 0\farcs09$, or 4.6~AU at 51 pc), we 
use the value $\Delta R = 0.5$~AU which \citet{cal02} find to be consistent 
with the shape of the mid-IR SED. We choose $R_0$ to be 4.5~AU so 
that the peak of the wall emission occurs near 4.0~AU, causing the 
null of the model to match the position indicated by the data.
For the dust mass opacity, we use the power law form
$\kappa = \kappa_0 (\frac{\nu}{\nu_0})^\beta $ where $\kappa_0 = 1.8$, 
$\nu_0 = 1.0\times 10^{12}$, and $\beta = 0.8$ \citep{dal01}.

We used the Monte Carlo radiative transfer code RATRAN \citep{hog00} 
to calculate a sky-projected image from the model continuum emission, 
with frequency and bandwidth appropriate for the observations,
and the miriad task {\em uvmodel} to simulate the observations,
using the appropriate antenna positions and visibility weights.
Figure~\ref{fig:vis} shows the visibility function from this model
(heavy solid line), which matches well the observations. 
The light solid line shows the contribution from the outer disk, which accounts 
for only a small fraction of the flux observed at long baselines,
and the dotted curve shows contribution from the thin wall component, 
which dominates the outer disk at ${\cal R}_{uv}$ beyond $\sim100$~k$\lambda$.
The long baseline data are sensitive to the total flux and thin width 
of the wall, and are compatible with the assumptions of the wall structure
used in the previous SED modeling. 
In the image domain, we have compared the 7~millimeter data to the \citet{cal02}
model using RATRAN to generate images of a disk with an inner hole of radius
4.5~AU and a bright wall of width 0.5~AU and total flux 1.7 mJy (Figure 
\ref{fig:map}, center panel), which \citet{cal02} find to be consistent with 
the mid-infrared SED.  For comparison, we also generate an image of a 
continuous disk extending inward to the dust destruction radius at 0.01~AU 
(Figure \ref{fig:map}, right panel).  
The data are inconsistent with this continuous disk model, while the SED-based
model of \citet{cal02} with an inner hole and bright wall matches the observed 
structure well.

In short, the 7~millimeter data suggest an outer dust disk extending inward 
to $\sim4.5$~AU, where there is wall of radial extent $\sim$0.5~AU that is 
much brighter than the surrounding disk on account of direct illumination by 
the star.  Interior to the wall, a sharp transition occurs to a region of 
lower dust surface density and correspondingly weak or absent 
7~millimeter emission.

\subsection{Disk Clearing}

The resolved observations bolster the inference from SED models that 
the TW Hya disk has an inner hole of much reduced dust column density 
of radius $\sim4$~AU. 
Theories of disk-planet interaction have long predicted the opening of 
gaps in circumstellar disks as a consequence of the formation of 
giant planets \citep[e.g.][]{lin86,bry99}, and numerical simulations of 
such interactions produce these gaps in a consistent way across differing 
architectures and computational algorithms \citep{val06}. 
Recent work by \citet{var06} has shown that in turbulent disks with 
$\alpha$-viscosity, inner holes are in fact more likely than gaps as a 
consequence of the formation of Jupiter-mass planets around solar mass stars, 
as the process of outward angular momentum transfer mediated by 
spiral density waves can cause clearing of the inner disk on time scales 
an order of magnitude shorter than the viscous timescale. 

While the inner disk is largely cleared, it is not entirely devoid of 
gas and dust.  Optical emission lines indicate gas accretion 
\citep{muz00}, albeit at the low rate of $4\times10^{-10}$~M$_\sun$~yr$^{-1}$. 
\citet{ret04} detect $\sim 6 \times 10^{21}$ g of warm CO at distances of 
0.5-1~AU from the star, and \citet{her04} infer the presence of warm H$_2$ 
within 2~AU of the star by modeling the HST and FUSE H$_2$ emission spectrum.
The 10~$\mu$m silicate feature \citep{sit00,uch04} indicates that there must be 
at least a few tenths of a lunar mass of dust grains present in the otherwise
largely cleared inner disk \citep{cal02}. 
The region within $\sim$0.3~AU has been spatially resolved by \citet{eis06} 
at 2~$\mu$m using the Keck interferometer, detecting emission from 
optically thin, submicron-sized dust populating the inner regions.
This small amount of material may be the result of a restricted flow 
through the transition region at $\sim4$~AU from the massive reservoir of 
the outer disk.
A population of micron-sized grains in the inner disk is consistent with
the predictions of \citet{ale06b}, who show that planet-induced gaps tend to 
filter out and confine large grains at the gap edge while allowing small
grains to migrate across the gap with accreting gas and populate the inner 
disk.  
Our data are consistent with an inner region devoid of 7mm dust emission; however, the degree of clearing is difficult to constrain due to both the low signal-to-noise ratio at long baselines and the dependence on the wall model adopted.  


The dynamical effect of a planet is not the only possible explanation 
for the observed central flux deficit at millimeter wavelengths. 
Photoevaporation of gas by ultraviolet radiation has also been invoked
to explain inner disk clearing \citep{cla01}, if the evaporation rate 
at the gravitational radius dominates the accretion rate. 
As discussed by \citet{ale06}, TW Hya is at best marginally consistent 
with a photoevaporation scenario, since the outer disk mass is larger than 
predicted for photoevaporation, and the observed accretion should occur 
only for the brief period while the inner disk is draining onto the star. 
In contrast with the case of a planet-induced gap, photoevaporation should also 
clear the inner disk of all but the largest grains, as noted by \citet{ale06b}.
Other mechanisms, such as photophoretic effects \citep{kra05}, could also 
aid in clearing the inner disk region, if gas densities are high enough. 

\section{Conclusions}

We present new, spatially resolved observations of the TW Hya disk at 
7~millimeters that provide direct evidence for a sharp transition 
in dust surface density at $\sim4$~AU radius, a feature consistent
with the inner hole inferred from SED modeling by \citet{cal02}.
The interpretation of the mid-infrared flux deficit 
as a central clearing of material is robust. The TW Hya system is ideally 
suited to future observations that will be able to distinguish between the 
various scenarios invoked to explain central clearing in the disks around 
young stars. Signatures diagnostic of planets in formation, in particular 
spiral density waves in the disk and thermal emission from circumplanetary 
dust, should be within the detection capabilities of the 
Atacama Large Millimeter Array operating at the shortest submillimeter 
wavelengths (300~$\mu$m) and at the longest baselines ($>10$~km) to achieve 
the necessary angular resolution \citep{wol05}. 
Such observations will benefit from the close proximity of TW Hya, 
the nearly face-on viewing geometry, and the size scale of the inner hole
now confirmed by direct observation at millimeter wavelengths.

\acknowledgements
Partial support for this work was provided by NASA Origins of 
Solar Systems Program Grant NAG5-11777.  MRH is supported by a VIDI
grant from the Netherlands Organization for Scientific Research (NWO). 

\appendix

\section{Protoplanetary Disk Visibility Functions}
\label{sec:viscurves}

We analyze a simplified disk model parameterized by power law distributions 
in surface density and temperature and including a central hole, to illustrate how
the model parameters affect the shape of the visibility function.
We also discuss the visibility function of a model consisting of a 
thin ring.  For both cases, we provide analytical expressions 
for calculating the position of the null in the (deprojected) visibility 
function, which is an easily observed feature.

\subsection{Power-Law Disk with a Central Hole}
\label{sec:pld}

For a flat, optically thin disk described by power-law distributions in 
temperature and surface density, assumed to be radiating in the 
Rayleigh-Jeans limit and viewed face-on, the intensity of radiation as a 
function of radial angular scale $\theta$ from the center of the disk is 
\begin{equation}
I_\nu(\theta) = \frac{2 \nu^2 k_B \kappa_\nu}{c^2} T_0 \left(\frac{\theta}{\theta_0}\right)^{-q} \Sigma_0 \left(\frac{\theta}{\theta_0}\right)^{-p}
\label{i}
\end{equation}
The visibility as a function of (u,v) distance, $\cal{R}$, is given by the 
Fourier transform of the intensity:
\begin{equation}
V({\cal R})=\frac{4 \pi \nu^2 k_B \kappa_\nu T_0 \Sigma_0 \theta_0^{p+q}}{c^2} 
\int_{\theta_a}^{\theta_b} \theta^{1-(p+q)} J_0(2 \pi \theta {\cal R})  d\theta
\label{v}
\end{equation}
where $J_0$ is a zeroeth order Bessel function.  The integral term in the above expression can be evaluated as
\begin{eqnarray}
\int_{\theta_a}^{\theta_b} \theta^{1-(p+q)} J_0(2 \pi \theta {\cal R})  d \theta =
\frac{\theta_b^{2-(p+q)} \Gamma\left(1-\frac{p+q}{2}\right)}{2 \left( \theta_b - \theta_a \right)} \left[ f(\theta_b) - f(\theta_a) \right] \equiv {\cal F}(\theta_b) - {\cal F}(\theta_a)
\label{integral}
\end{eqnarray}
where
\begin{eqnarray}
f(\theta) = \frac{1}{\Gamma\left(2-\frac{p+q}{2}\right)} ~_1F_2 \left[\{1-\frac{p+q}{2}\},\{1,2-\frac{p+q}{2}\},-\pi^2 {\cal R} ^2 \theta ^2 \right] 
\end{eqnarray}
and $_xF_y(\mathbf{a},\mathbf{b},z)$ is a generalized hypergeometric function of one variable.
For a disk with an inner hole, the limits of integration $\theta_a$ and 
$\theta_b$ are the inner and outer angular radii of the disk, respectively.   

For characterizing the visibility function on 
angular scales $\frac{1}{{\cal R}}$ between $\theta_a$ and $\theta_b$, we can 
make the approximation ${\cal R}^2 \theta_a^2 \ll 1$ and ${\cal R}^2 \theta_b^2 \gg 1$.  
In these limits, the hypergeometric function attains manageable analytic forms. 

\noindent
For $z \ll 1$, appropriate for the inner disk of radius $\theta_a$, the function $_xF_y(\mathbf{a},\mathbf{b},z)$ has the following series expansion:
\begin{eqnarray}
\lim_{z\to 0}~_xF_y(\mathbf{a},\mathbf{b},z)= \sum_{k=0}^{\infty} \frac{ \prod_{i=1}^x PH(a_i,k) }{ \prod_{j=1}^y PH(b_j,k) } \frac{z^k}{k!} \nonumber  \\
PH(c,k)= (c+k)!/c!
\label{zlt1}
\end{eqnarray}
and in this limit the quantity ${\cal F}$ goes to 
\begin{eqnarray}
{\cal F}(\theta_a) = \frac{\theta_a^{2-(p+q)}}{2-(p+q)} \lim_{{\cal R}\theta_a \to 0}~_1F_2 \left[\{1-\frac{p+q}{2}\},\{1,2-\frac{p+q}{2}\},-\pi^2 {\cal R} ^2 \theta_a^2 \right]
\label{thetaa}
\end{eqnarray}
with the limit approximated by the sum in \ref{zlt1} above.

\noindent
In the limit of $z \gg 1$, appropriate for the outer radius ${\cal R}^2 \theta_b^2$, 
the quantity ${\cal F}$ approaches the analytical form
\begin{eqnarray}
\lim_{{\cal R}\theta_b \to \infty}~{\cal F}(\theta_b)=\frac{ |\pi {\cal R} \theta_b|^{p+q-2} \Gamma(2-\frac{p+q}{2}) }{ \Gamma(\frac{p+q}{2}) }
\label{zgt1}
\end{eqnarray}

\noindent
The total visibility function (\ref{v}, with integral \ref{integral}) is then the difference 
between a smooth power-law disk without a hole (${\cal F}(\theta_b)$, equation \ref{zgt1}) and the contribution of the evacuated inner region (${\cal F}(\theta_a)$, equation \ref{thetaa} with limit \ref{zlt1}).

\subsubsection{Position of the Null}

\begin{figure}
\centering
\epsscale{0.6}
\plotone{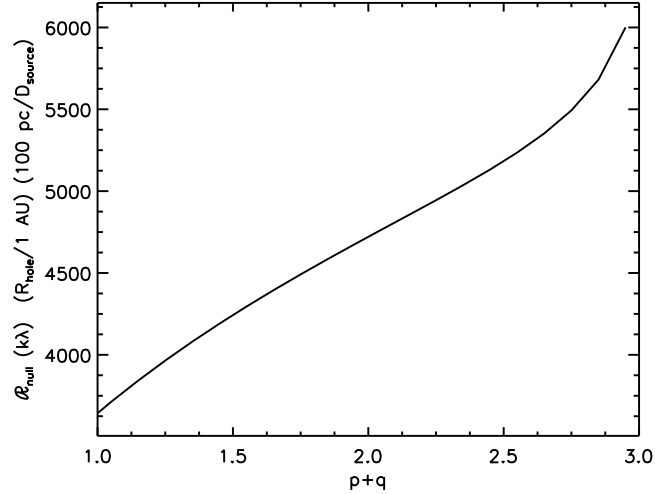}
\caption{The dependence of ${\cal R}$ on disk temperature and surface 
density power law indices ($p+q$), source distance ($D_{source}$), and inner hole size ($R_{hole}$).  
}
\label{fig:p+q}
\end{figure}

\begin{figure}[t]
\centering
\epsscale{0.6}
\plotone{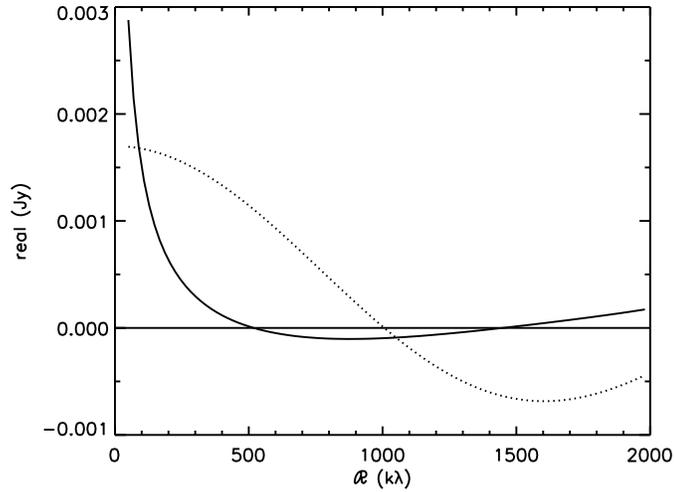}
\caption{
The visibility function for the TW Hya disk, based on the power-law model 
with an inner hole in which expansion \ref{zlt1} is carried to third order (solid line). 
The visibility function for a thin ring interior to the 
inner disk edge is also shown (dotted line; see equation \ref{eq:wall}).
}
\label{fig:viscurve}
\end{figure}

For a disk with an inner hole, the position of a null in the visibility
function is an easily observed quantity. Here we show how the angular scale 
of the first null depends on the disk model parameters.
Substituting expansion \ref{zlt1} and equation \ref{zgt1} into \ref{integral} 
and setting the result equal to zero, we obtain the following expression 
which can be solved for the position of the first null:
\begin{eqnarray}
\frac{\Gamma \left (1-\frac{p+q}{2} \right) \left(1-\frac{p+q}{2}\right) }{\Gamma \left(\frac{p+q}{2} \right )  } = (\pi {\cal R}_{null} \theta_a)^{2-(p+q)} \left( 1-\frac{ 2-(p+q) }{ 4-(p+q) } \pi^2 {\cal R}_{null}^2 \theta_a^2 + \cdots \right)
\label{null1}
\end{eqnarray}
Since ${\cal R}$ and $\theta_a$ always appear in tandem in this expression, 
${\cal R}_{null} \propto \frac{1}{\theta_a}$.

The dependence of the null position on the power law indices $p$ and $q$ is illustrated in Figure~\ref{fig:p+q}, which shows the null position as a function of $p+q$ for a fixed inner disk radius.  
Several orders of the power series expansion (\ref{zlt1}) are shown.
For typical values 
of the power law indices, the null position shifts monotonically to longer 
baselines as $p+q$ increases, exhibiting an essentially linear relationship  
in the vicinity of $p+q$=2.  
As $p+q$ increases, the temperature and surface density distributions (and 
therefore the intensity) become more sharply centrally peaked, and so the 
position of the null, which is effectively the angular scale on which the 
inner disk contribution to the visibility equals that of the outer disk, 
moves to smaller and smaller angular scales (i.e., larger ${\cal R}$).

These approximations begin to lose validity longwards of the vicinity of the 
first null, which occurs at ${\cal R} \theta_a < 0.3$ for all $p+q < 3$.  
The series expansion quickly diverges past ${\cal R} \theta_a = 1$.  
However, it should be also noted that this constraint places no limit on the 
size of the hole that can be investigated by this method, i.e., for any disk 
with a central hole there will always be at least one null shortward of 
${\cal R} \theta_a$=1, and so this method is robust for any case in which $\theta_b$ 
is large compared to $\theta_a$. 
It is also valid for an inclined (symmetric) disk, as long as the 
deprojection is handled appropriately, as in \S\ref{sec:deproj}. 

For ease of use, it is possible to approximate the $(p+q)$ dependence by a linear fit of the curve in the region $1 \le p+q \le 3$, which results in the following formula for the position of the null, good to within 4\%:
\begin{eqnarray}
{\cal R}_{null} (k\lambda) =\left ( \frac{1~AU}{R_{hole}} \right ) \left ( \frac{D_{source}}{100~pc} \right ) \left [ 2618 + 1059 \times (p+q) \right] 
\label{eq:approx}
\end{eqnarray}

\subsection{Thin Wall}
\label{sec:thwall}

A thin wall ($\Delta R \ll$ R) 
can be described by a ring of constant brightness at a distance $R_{hole}$ 
from the star, with a visibility function which is a zeroeth order 
Bessel function:
\begin{eqnarray}
V({\cal R})=2 \pi \int_{0}^{\infty} I_\nu(\theta) J_0(2 \pi \theta {\cal R}) \theta d\theta \nonumber \\
= 2 \pi \theta_{hole} I_{wall} J_0(2 \pi \theta_{hole} {\cal R})
\label{eq:wall}
\end{eqnarray}
where $\theta_{hole}$ is the angular radius of the hole and $I_{wall}$ is the 
intensity of emission from the wall. 
The position of the null in the visibility function of a thin wall will then be 
\begin{eqnarray}
{\cal R}_{null}(k \lambda) = \frac{77916}{\pi^2} \left( \frac{D_{source}}{100~pc} \right) \left( \frac{1~AU}{R_{hole}} \right)
\label{eq:bessel}
\end{eqnarray}

\subsection{Application to TW Hya}

The TW Hya disk has an outer radius of 196~AU (CO emission,
\citet{qi04}) and an inner hole with radius 4~AU 
(SED models, \citet{cal02}; imaging, this paper).
These size scales correspond to (u,v) distances of 
54 k$\lambda \le {\cal R} \le$ 2600 k$\lambda$,
a range well matched to the coverage of the Very Large Array 
7~millimeter observations, and we may apply the method 
of \S\ref{sec:pld} to generate the visibility function 
based on the simple power-law disk model.
Approximating the \citet{cal02} model 
with power laws in surface density 
and temperature yields the profile 
  $T(R)=T_{100} \left(R/100~AU\right)^{-q}$ 
and 
  $\Sigma(R)=\Sigma_{100} \left(R/100~AU\right)^{-p}$ 
where $T_{100}$= 28 K, $q$=0.44, $\Sigma_{100}$=3.7$g~cm^{-2}$, 
and $p$=0.90
For the dust opacity, we adopt a power-law distribution with 
  $\kappa_\nu = \kappa_0 \left ( \nu/\nu_0 \right )^\beta$, 
where $\kappa_0 = 1.8 cm^2~g^{-1}$, $\nu_0 = 1.0 \times 10^{12}$Hz, 
and $\beta = 0.8$ \citep{dal01}.  
Figure \ref{fig:viscurve} shows the visibility function
calculated according to equation \ref{v} with the limits as in 
equations \ref{zlt1} and \ref{zgt1}, expanding to third order 
in $({\cal R} \theta_a)^2$.  
The resulting curve agrees well with the visibility function derived
for the power law disk model obtained with the full radiative transfer 
calculation, shown in Figure~\ref{fig:vis} (light solid line).
Figure~\ref{fig:viscurve} also shows the visibility function of a thin ring 
interior to the disk (dotted line), as in \S\ref{sec:thwall}, with radius 
4~AU and flux 1.7 mJy.

\end{document}